\documentclass{icsm}
\usepackage{amsmath,amssymb}
\usepackage{pstricks}

\slacs{.4ex}

\newcommand{\pd}{\partial}
\newcommand{\ket}[1]{\left|#1\right\rangle}
\newcommand{\bracket}[2]{\left\langle%
#1\left.\right|#2\right\rangle}
\newcommand{\tr}{\mathop{\mathrm{tr}}\nolimits}
\newcommand{\Op}{\mathcal{O}}

\begin{document}
\title{Chaining spins from (super)Yang\bmth{-}Mills}
\authori{S. Bellucci, P.Y. Casteill, J.F. Morales,
C. Sochichiu\footnote{On leave from: Bogoliubov Lab. Theor. Phys.,
JINR, 141980 Dubna, Moscow Reg., RUSSIA and Institutul de Fizic\u
a Aplicat\u a A\c S, str. Academiei, nr. 5, Chi\c sin\u au, MD2028
MOLDOVA.}}
\addressi{INFN -- Laboratori Nazionali di Frascati,
Via E. Fermi 40, 00044 Frascati, Italy}
\authorii{}     \addressii{}
\authoriii{}    \addressiii{}
\authoriv{}     \addressiv{}
\authorv{}      \addressv{}
\authorvi{}     \addressvi{}
%
\headauthor{S. Bellucci, P.Y. Casteill, J.F. Morales, C.
Sochichiu}
\headtitle{Chaining spins from (super)Yang--Mills}
\lastevenhead{S. Bellucci et al.: Chaining spins from
(super)Yang--Mills}

\maketitle

\begin{abstract}
We review the spin bit model describing anomalous dimensions of
the operators of Super Yang--Mills theory. We concentrate here on
the scalar sector. In the limit of large $N$ this model coincides
with integrable spin chain while at finite $N$ it has nontrivial
chain splitting and joining interaction.
\end{abstract}

\section{Introduction}

Effective description of gauge interactions in terms of string has a long history
\cite{'tHooft:1974jz} (see \cite{'tHooft:2002yn} for a more recent review). Last
years the development of the field has streamlined into what is know as AdS/CFT
(or Maldacena) conjecture \cite{Maldacena:1998re,Gubser:1998bc}, which claims that
superstring theory in the background of AdS$_5\times S^5$ is essentially the same
as $\mathcal{N}=4$ Super Yang--Mills (SYM) model in four--dimensional Minkowski
space which is the topological boundary of AdS$_5$. The assumption is based on the
fact that two theories have the same symmetry group whose bosonic part
$SO(2,4)\times SU(4)$ is one one hand the symmetry group of AdS$_5\times S^5$
space and one the other hand is the conformal group of four--dimensional Minkowski
space. (Due to vanishing of $\beta$--functions $\mathcal{N}=4$ SYM model is a
conformally invariant model.)

Identifying irreducible representation of both groups one can put into one-to-one
correspondence operators in SYM and states in the string theory. In particular,
anomalous dimensions of operators in SYM theory (which are eigenvalues of the
dilations) correspond to energy levels in string theory (see \cite{Aharony:1999ti}
for a review of AdS/CFT correspondence).

Recently much progress was achieved in understanding the scaling
properties of SYM operators (see e.g. \cite{Plefka:2003nb}). In
particular, it was found for the scalar sector of SYM theory
operators that in the planar limit the anomalous dimension matrix
can be mapped into the Hamiltonian of integrable $SU(4)\sim SO(6)$
spin chain \cite{Minahan:2002ve}, this result was further
generalized to the whole symmetry supergroup $SU(2,2|4)$ in
\cite{Beisert:2003yb}. Thus, Bethe Ansatz allows one to find the
planar anomalous dimension of any operator in SYM without
computing explicitly the corresponding Feynman diagrams. This spin
chain is supposed to be a discrete version of the string in AdS
background.

Going beyond the planar limit results in allowing chains to split and join. (When
a fixed number of impurities is considered this dynamics can be described in terms
of a quantum mechanical system like one in \cite{Beisert:2002ff}.) The natural
task is, then, to extend the spin chain description to the nonplanarity. Indeed,
one can do a one-to-one map of SYM operators into a system of interacting spins
--- \emph{spin bits} such that the anomalous dimension matrix of SYM operators
maps to a Hamiltonian for this spin system at finite $N$. In this note we are
going to review this model. Interesting reader can find more details in the
original paper \cite{Bellucci:2004ru}.

\section{Preliminaries}

We consider scalar $k$--trace operators of the type\footnote{The
number of traces can vary, it is a dynamical quantity},
\begin{equation}\label{ops}
  \Op=\tr
  \phi_{i^{(1)}_1}\cdots\phi_{i^{(1)}_{L_1}}\tr\phi_{i^{(2)}_1}
  \cdots\phi_{i^{(2)}_{L_1}}\dots
  \tr\phi_{i^{(k)}_1}\cdots\phi_{i^{(k)}_{L_k}}\,,
\end{equation}
where $L_i$ are the lengths of traces and $L=L_1+L_2+\dots+L_k$ is
the total length. The above operators can be equivalently
represented in the following form,
\begin{equation}\label{ops-gamma}
  \Op\equiv\ket{s;\gamma}=
  \phi^{a_1a_{\gamma(1)}}_{i_1}\phi^{a_2a_{\gamma(2)}}_{i_2}\dots
  \phi^{a_La_{\gamma(L)}}_{i_L},
\end{equation}
where $\gamma$ is an element of the permutation group of
$\{1,2,\dots,L\}$ and $s$ labels the indices $i_1,i_2,\dots,i_L$.
In particular, the operator \eqref{ops} corresponds to a
permutation with cycles $(L_1)(L_2)\dots(L_k)$. It is not
difficult to see that to each permutation one can put into
correspondence a trace of operators if one specifies the index of
each letter.

Graphically the field $\phi$ insertion will be represented by a
site with two valent lines, like follows:
\begin{center}
  \input{soch1.pic}
\end{center}
The incoming arrow connects with the site $\gamma^{-1}(k)$ while
the outgoing one goes to $\gamma(k)$. In general, the arrow denote
the action of the permutation.

\section{The Hamiltonian}

The Hamiltonian in the combinatorial form is obtained by computing
the action of the operator \cite{Beisert:2002bb}
\begin{equation}\label{beisert}
  H\equiv\Delta_{(2)}=-:\tr [\phi_m,\phi_n][\check{\phi}_m,\check{\phi}_n]:-
  \sfrac{1}{2}:\tr[\phi_m,\check{\phi}_n][\phi_m,\check{\phi}_n]:\,,
\end{equation}
where,
\begin{equation}\label{checks}
  \check{\phi}^{ab}_i=\frac{\pd}{\pd \phi^{ba}_i}\,,
\end{equation}
and colons denote that derivatives do not act on other fields in the group.

Let us compute directly the action of the operator \eqref{beisert}
on a state $\ket{s;\gamma}$, \be\label{H-state:1}\ba{rcl}
H\ket{s;\gamma}&=&
\Bigl[2(\delta^{j_1j_4}\delta^{j_2j_3}-\delta^{j_1j_3}\delta^{j_2j_4})
:\tr\phi_{j_1}\phi_{j_2}\check{\phi}_{j_3}\check{\phi}_{j_4}:+\\[4pt]
&&\quad+\delta^{j_1j_3}\delta^{j_2j_4}
:\tr\phi_{j_1}\check{\phi}_{j_2}\phi_{j_3}\check{\phi}_{j_4}:+\\[4pt]
&&\qquad+\delta^{j_1j_4}\delta^{j_2j_3}
:\tr\phi_{j_1}\check{\phi}_{j_2}\check{\phi}_{j_3}\phi_{j_4}:\Bigr]
\bigl(\phi^{a_1a_{\gamma(1)}}_{i_1}\dots
\phi^{a_La_{\gamma(L)}}_{i_L}\bigr). \ea\ee

\begin{figure}[t]
  \centering{\input{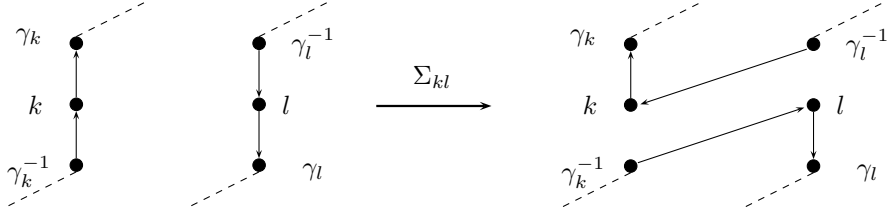}}
\vskip-10mm
  \caption{Action of $\Sigma_{kl}$ on the permutation state $\gamma$.}\label{fig:sigma}
\end{figure}

Application of the operator from the second line of
\eqref{H-state:1} yields, \be\label{second-line}\ba{l}
2\delta^{a_ka_{\gamma(l)}}\left(\phi^{a_1a_{\gamma(1)}}_{i_1}\dots
  \phi^{a a_{\gamma(k)}}_{i_k}\dots\phi^{a_la}_{i_l}
  \dots\phi^{a_La_{\gamma(L)}}_{i_L}\right)-\\[6pt]
\qquad\qquad\qquad
-2\delta^{a_ka_{\gamma(l)}}\left(\phi^{a_1a_{\gamma(1)}}_{i_1}\dots
  \phi^{a_l a}_{i_k}\dots\phi^{aa_{\gamma(k)}}_{i_l}
  \dots\phi^{a_La_{\gamma(L)}}_{i_L}\right).\ea\ee
This corresponds to graphs with modified cyclic structure:
$\gamma\mapsto \gamma\cdot\sigma_{k\gamma_l}$ and
$\gamma\mapsto\sigma_{kl}\cdot\sigma_{k\gamma_l}\cdot\gamma\cdot\sigma_{kl}$
respectively for the first and second line of \eqref{second-line}
(see fig. \ref{fig:sigma}. Therefore, action of this part of the
Hamiltonian \eqref{beisert} can be represented as,
\begin{equation}\label{2nd-line}
2\bigl[\ket{s;\gamma\cdot\sigma_{k\gamma_l}}-
\ket{s;\sigma_{kl}\cdot\gamma\cdot\sigma_{k\gamma_l}\cdot\sigma_{kl}}\bigr].
\end{equation}

Analogously, the second line of \eqref{H-state:1}  produce the  modified cycles
given by,
\begin{equation}\label{3d-line}
\sum_{kl}K_{kl}\bigl[\ket{s;\gamma\cdot\sigma_{k\gamma_l}}-
  \ket{s;\gamma\cdot\sigma_{\gamma_k\gamma_l}}\bigr]\,.
\end{equation}

Combining both results \eqref{2nd-line} and \eqref{3d-line}
together, one gets for the total Hamiltonian~:
\be\label{hamiltonian}\ba{rcl} H\ket{s;\gamma}&=&\disty
 \sum_{kl}2\bigl(\ket{s;\gamma\cdot\sigma_{k\gamma_l}}-
  \ket{s;\sigma_{kl}\cdot\gamma\cdot\sigma_{k\gamma_l}\cdot\sigma_{kl}}
  \bigr)+\\[12pt]
&&\disty+\sum_{kl}K_{kl}\bigl(\ket{s\gamma\cdot\sigma_{k\gamma_l}}-
  \ket{s;\gamma\cdot\sigma_{\gamma_k\gamma_l}}\bigr)=\\[12pt]
&=&\disty\sum_{kl}\bigl[2(1-P_{k\gamma^{-1}(l)})
  +(K_{k\gamma^{-1}(l)}-K_{\gamma^{-1}(k)\gamma^{-1}(l)})\bigr]
  \ket{s\gamma\cdot\sigma_{kl}}\equiv\\[12pt]
&\equiv&\disty
  \sum_{kl}(H_{k\gamma^{-1}(l)}-H_{\gamma^{-1}(k)\gamma^{-1}(l)})\Sigma_{kl}
  \ket{s;\gamma}.\ea\ee
Here we introduced the joining/splitting operator $\Sigma_{kl}$,
which change the permutation group element as follows,
\begin{equation}\label{gamma}
  \Sigma_{kl}\ket{s;\gamma}=
  \begin{cases}
    \ket{s;\gamma\cdot\sigma_{kl}},& k\neq l\,,\\
    N\ket{s;\gamma},& k=l\,.
  \end{cases}
\end{equation}
Factor $N$ for coinciding $k$ and $l$ appears due to the fact that
splitting and joining the trace in the same point results in
multiplying by $\tr\mathrm{Id}=N$.

Note that since $\gamma_k$ appears as index in the spin part
\eqref{hamiltonian} the order between $\Gamma_{kl}$ and $H_{kl}$
is important. One can, however re-sum \eqref{hamiltonian} in order
to get a form in which spin and permutation parts are completely
independent and commute:
\begin{equation}\label{spin-ham}
H=\sum_{k,l}H_{kl}(\Sigma_{k\gamma_l}-\Sigma_{\gamma_k\gamma_l})\,.
\end{equation}

\section{The Hilbert space}

Consider the description of the operators \eqref{ops-gamma} in
terms of the spin states. The $L$-field operator like \eqref{ops}
should be labelled by the spin variables $\{s^{\alpha}_n\}$ and a
permutation group element $\gamma$.

The spin part itself is represented by a vector in the tensor
product of one spin representations corresponding to each ``site''
$n$,
\begin{equation}
  \ket{s_1}\otimes\ket{s_2}\dots\otimes\ket{s_L}\equiv\ket{s}.
\end{equation}

Obviously, one can choose an orthonormal, basis in the one--spin
space such that, $s_n=s^\alpha_n e_\alpha$ and
\begin{equation}
  \bracket{e_{\alpha}}{e_{\beta}}=\delta_{\alpha\beta}\,.
\end{equation}

The ``graph'' part of the state we denote by $\ket{\gamma}$,
$\gamma\in \Gamma_L$, permutation group of $L$ elements. For the
space of permutation there is a natural scalar product,
\begin{equation}\label{gamma-scalar}
\bracket{\gamma}{\gamma'}=\delta_{\gamma\gamma'}\,,
\end{equation}
i.e. it is zero for different permutations and unity when
contracted with itself.

The states in our model should be represented as elements tensor
product of the above two spaces modulo the symmetry group $S_L$,
\begin{equation}\label{state}
  \ket{\{s\},\gamma}\in\{\ket{\{s\}}
  \otimes\ket{\gamma}\}/S_L\,,
\end{equation}
where the symmetry group acts as follows,
\begin{equation}
  \hat{\Sigma}_\sigma\ket{s}\otimes\ket{\gamma}=\ket{s_{\sigma}}
  \otimes\ket{\sigma^{-1}\cdot\gamma\cdot\sigma},\quad
  \sigma\in S_L\,,
\end{equation}
where
\begin{equation}
  s_{\sigma}=\{s_{\sigma(1)},s_{\sigma(2)},\dots,s_{\sigma(L)}\}
\end{equation}
is a permutation of indices given by $\sigma\in S_L$. Indeed, as
it is not very difficult to see the original and permuted states
describe, in fact, the same trace in SYM model.

Given an arbitrary basis element $\ket{s}\otimes\ket{\gamma}$ one
can find an element of the factor \eqref{state} by averaging with
respect to the action of the group $\Gamma_L$,
\begin{equation}\label{ph-state}
  \ket{s;\gamma}=\frac{1}{|S_L|}\sum_{\sigma\in S_L}\ket{s_{\sigma}}
  \otimes\ket{\sigma^{-1}\cdot\gamma\cdot\sigma}\equiv \hat{\Pi} \ket{s}
  \otimes\ket{\gamma},
\end{equation}
where $\Pi$ is the cyclic symmetry projector,
\begin{equation}\label{Pi}
\Pi=\frac{1}{|S_L|}\sum_{\sigma\in
S_L}U_{\sigma}\otimes\hat{\Sigma}_{\sigma}\,,
\end{equation}
where $U_{\sigma}$ and $\hat{\Sigma}_{\sigma}$ are the following
operators
\begin{equation}
U_{\sigma}=P_{1\sigma(1)}\otimes P_{2\sigma(2)}\dots\otimes
P_{L\sigma(L)}\,,\quad
\hat{\Sigma}_{\sigma}\ket{\gamma}=\ket{\sigma^{-1}\cdot\gamma\cdot\sigma}
\end{equation}
and $|S_L|$ stays for the order of $S_L$. As we mentioned $\Pi$ is
a projector, i.e. $\Pi^2=\Pi$, also it is not difficult to see
that $\Pi$ commutes with permutation invariant operators.

Obviously, the above states are invariant with respect to the
action of the gauge group $S_L$. In particular, when
$\sigma=\gamma$ this symmetry represents the cyclicity of the
trace(s).

\section{Gauge symmetry}

Let us show that the spin bit Hamiltonian \eqref{spin-ham} can be
seen as arising from gauging of the planar spin chain. Since
$\gamma$ has the natural interpretation of the connection then the
$\gamma$ preserving symmetry $n\mapsto\sigma(n)$ has the meaning
of the ``global'' gauge symmetry. (In fact this is translation
symmetry.)

As we have seen in the previous section, an arbitrary permutation
leads to more general transformation rules for ``points'' and
``connections''
\begin{equation}\label{transf}
   n\mapsto\sigma(n)\,,\quad
   \gamma\mapsto\sigma^{-1}\cdot\gamma\cdot\sigma\,.
\end{equation}
Now, this is the localized version of the shift transformation,
i.e. the discrete analog of the diffeomorphism transformations.

By direct evaluation of the Hamiltonian \eqref{spin-ham} one can
show that it is invariant with respect to transformations
\eqref{transf}. The Hamiltonian \eqref{spin-ham} can be rewritten
in the following form,
\begin{equation}\label{h-gauged}
  H=\sum_{kl}V_{kl}H_{kl}\,,
\end{equation}
where,
\begin{equation}\label{v}
  V_{kl}=\Sigma_{k\gamma_l}-\Sigma_{\gamma_k\gamma_l}\,.
\end{equation}
In above expression $V_{kl}$ can be expressed as the discrete
gauge connection between sites $k$ and $l$.

On the other hand, consider the planar spin chain Hamiltonian,
\begin{equation}\label{h-planar}
  H_{0}=\sum_k H_{k,k+1}=\sum_{kl}V^{(0)}_{kl}H_{kl}\,,
\end{equation}
where,
\begin{equation}\label{v0}
V^{(0)}_{kl}=\delta_{k\gamma_0(l)}-\delta_{\gamma_0(k)\gamma_0(l)}\,,\quad
  \gamma_0(n)=n+1\,.
\end{equation}
The above expression \eqref{v0} differs from the complete
nonplanar $V_{kl}$ in eq. \eqref{v} by the only fact that
operators $\Sigma_{kl}$ are replaced by Kronecker delta symbols
$\delta_{kl}$.

Now it is clear, that passing from planar to general non-planar description
amounts in ``switching on'' the gauge field operator $V_{kl}$ (or $\Sigma_{kl}$), which
plays the role of the gauge field. This procedure is sensitive to transformation
properties of the fields to the permutation group action only and not on the
structure of $H_{kl}$. Therefore, this procedure can be applied for obtaining the
nonplanar Hamiltonian in the case of the spin chain with arbitrary group just
passing from $V^{(0)}$ to $V$.

\bigskip\noindent
{\small\textbf{Acknowledgements.} C.S. thanks the organizers of
the workshop for kind hospitality. This work was partially
supported by NATO Collaborative Linkage Grant PST.CLG. 97938,
INTAS-00-00254 grant, RF Presidential grants MD-252.2003.02,
NS-1252.2003.2, INTAS grant 03-51-6346, RFBR-DFG grant 436 RYS
113/669/0-2, RFBR grant 03-02-16193 and the European Community's
Human Potential Programme under contract HPRN-CT-2000-00131
Quantum Spacetime.}

\providecommand{\href}[2]{#2}\begingroup\raggedright\endgroup

\end{document}